\documentclass[reprint,amssymb,amsmath,aip,cha]{revtex4-2}
\usepackage{bm}%
\usepackage{color}
\usepackage{graphicx}
\usepackage{mathptmx}
\usepackage{amsmath}
\usepackage{derivative}
\usepackage[utf8]{inputenc}
\usepackage[T1]{fontenc}
\usepackage{upgreek}

\DeclareUnicodeCharacter{2009}{\,} 

\begin{document}

\title{Nonlinear dynamics in magnonic Fabry-P\'{e}rot resonators: Low-power neuron-like activation and transmission suppression}

\author{Anton Lutsenko}
\affiliation{NanoSpin, Department of Applied Physics, Aalto University School of Science, PO Box 15100, FI-00076 Aalto, Finland}

\author{Kevin G. Fripp}
\affiliation{University of Exeter, Stocker Road, Exeter, EX4 4QL, United Kingdom}

\author{Luk\'{a}\v{s} Flaj\v{s}man}
\affiliation{NanoSpin, Department of Applied Physics, Aalto University School of Science, PO Box 15100, FI-00076 Aalto, Finland}

\author{Andrey V. Shytov}
\author{Volodymyr V. Kruglyak}\email[Corresponding author: ]
{v.v.kruglyak@exeter.ac.uk}
\affiliation{University of Exeter, Stocker Road, Exeter, EX4 4QL, United Kingdom}

\author{Sebastiaan van Dijken}
\email[Corresponding author: ]
{sebastiaan.van.dijken@aalto.fi}
\affiliation{NanoSpin, Department of Applied Physics, Aalto University School of Science, PO Box 15100, FI-00076 Aalto, Finland}


\begin{abstract}
We report on nonlinear spin-wave dynamics in magnonic Fabry–Pérot resonators composed of yttrium iron garnet (YIG) films coupled to CoFeB nanostripes. Using super-Nyquist sampling magneto-optical Kerr effect microscopy and micromagnetic simulations, we observe a systematic downshift of the spin-wave transmission gaps as the excitation power increases. This nonlinear behavior occurs at low power levels, reduced by a strong spatial concentration of spin waves within the resonator. The resulting power-dependent transmission enables neuron-like activation behavior and frequency-selective nonlinear spin-wave absorption. Our results highlight magnonic Fabry–Pérot resonators as compact low-power nonlinear elements for neuromorphic magnonic computing architectures.
\end{abstract}

\maketitle

Magnonics has emerged as a promising platform for designing energy-efficient and scalable devices for information processing\cite{Chumak2022}. Many existing magnonic devices and proof-of-concept demonstrations operate in the linear regime, where spin waves propagate and interfere without mutual interaction. While this regime is well understood and technologically valuable, the most versatile and powerful magnonic functionalities arise from nonlinear dynamics. Accessing the nonlinear regime enables a wide range of magnonic phenomena, including amplitude-dependent frequency shifts\cite{Suhl1960,Fetisov2002,Gui2009,Wang2020}, multi-magnon scattering\cite{Suhl1957,Cherepanov1993,Ordonez2009,Schultheiss2012,Korber2020}, parametric amplification\cite{Bagada1997,Kolodin1998}, formation of solitons and other self-localized modes\cite{Bagada1997,Kolodin1998,Slavin2002,Wu2006,Chen2016}, and generation of coherent spin-wave frequency combs\cite{Wang2021,Hula2022,Wang2024}. These effects form the basis for various applications, such as reconfigurable on-chip signal processing, microwave generation and control, neuromorphic computing\cite{Papp2021,Watt2021,Korber2023}, and quantum transduction\cite{Lauk2020,Bejarano2024,Puel2025}.

The inherent nonlinearity of magnetization dynamics manifests itself in a strong magnon-magnon interaction, whereas external stimuli\textemdash such as microwave magnetic fields, spin-transfer torque, and spin-orbit torque\textemdash can drive strongly nonlinear responses, enabling controlled synchronization and conversion of spin-wave modes. Geometric confinement and resonant enhancement can significantly strengthen these nonlinear effects further. For example, structures such as magnonic cavities\cite{Neuman2020,Santos2023}, ring resonators\cite{Wang2020}, and chiral resonators\cite{Kruglyak2021,Fripp2023,Fripp2025} reshape the mode spectrum and concentrate spin-wave energy, thereby lowering the power threshold for a magnonic device to enter the nonlinear regime. These engineered resonant environments also allow tuning of the power dependence of spin-wave transmission, supporting threshold-activation behavior analogous to biological and artificial neurons. Neuron-like functionality has been modeled in ring resonators\cite{Wang2020} and chiral magnonic resonators\cite{Fripp2023,Fripp2025}, and demonstrated experimentally using nonlinear excitation with coplanar waveguides in Ga-doped YIG films\cite{Breitbach2025}.

The recently introduced magnonic Fabry-P\'{e}rot resonator\cite{Qin2021b,Lutsenko2025} offers additional potential as a scalable nonlinear element. This structure consists of a low-damping yttrium iron garnet (YIG) film coupled to a ferromagnetic-metal nanostripe. Dynamic dipolar coupling between the two materials forms a magnonic cavity with a modified spin-wave dispersion. The cavity resonances lead to narrow gaps in the spin-wave transmission spectrum while maintaining low loss at intermediate frequencies. This enables field-tunable control of the spin-wave amplitude\cite{Qin2021b,Talapatra2023} and phase\cite{Lutsenko2025}. The strong wavelength reduction within the resonator also enables substantial miniaturization compared to other resonator concepts, paving the way toward dense networks of magnonic elements.   

\begin{figure}[b]
	\centering
	\includegraphics[width=\columnwidth]{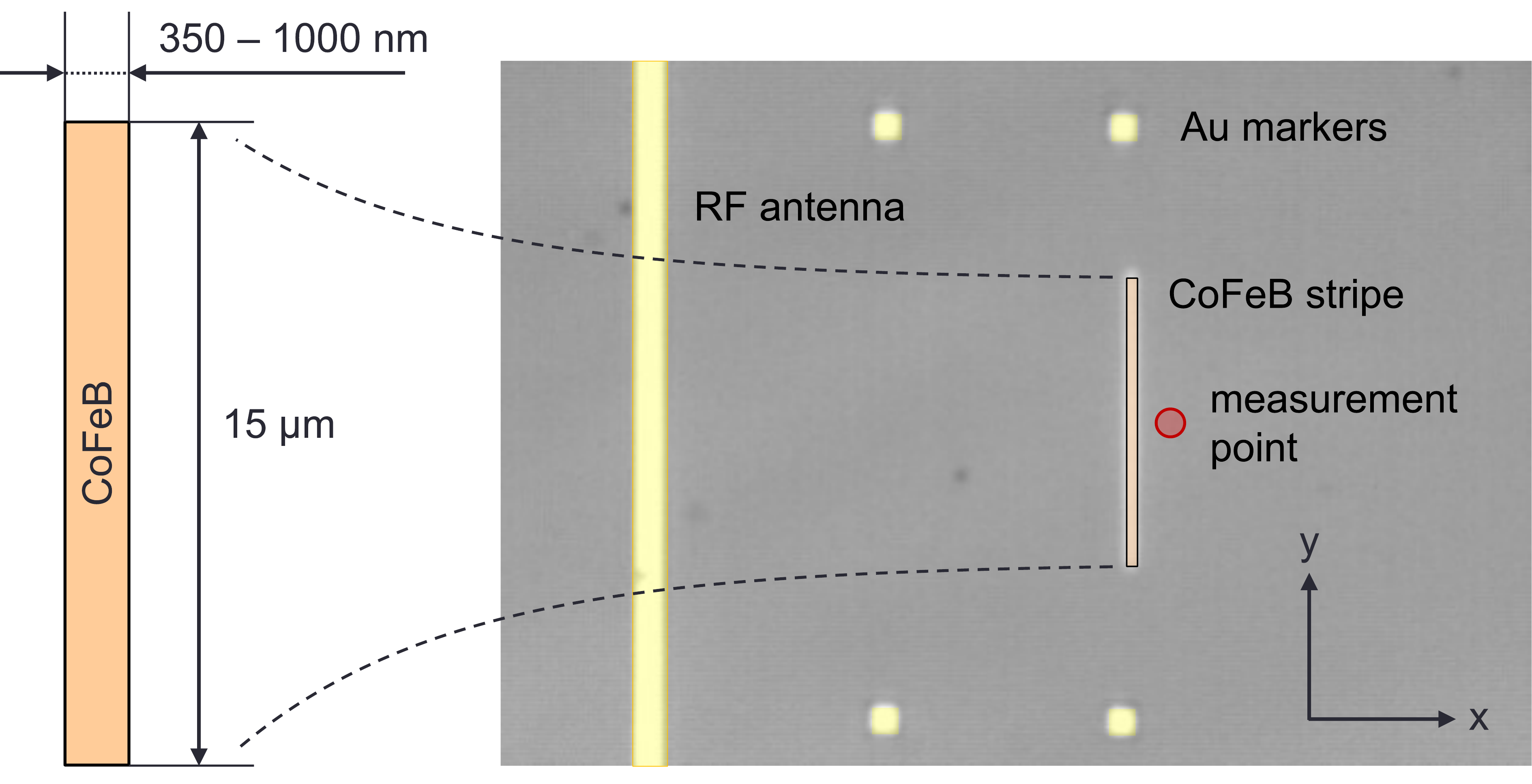}
	\caption{Microscopy image of the sample and the measurement geometry. The Fabry-P\'{e}rot resonator, consisting of a 30-nm-thick CoFeB nanostripe patterned on an 85-nm-thick YIG film, is positioned 25 $\upmu$m from a 1.5-$\upmu$m-wide microwave antenna. SNS-MOKE measurements were performed 2 $\upmu$m behind the resonator center, as indicated by the red circle.}
    \label{Fig1}
\end{figure}

\begin{figure*}[t]
	\centering
	\includegraphics[width=2\columnwidth]{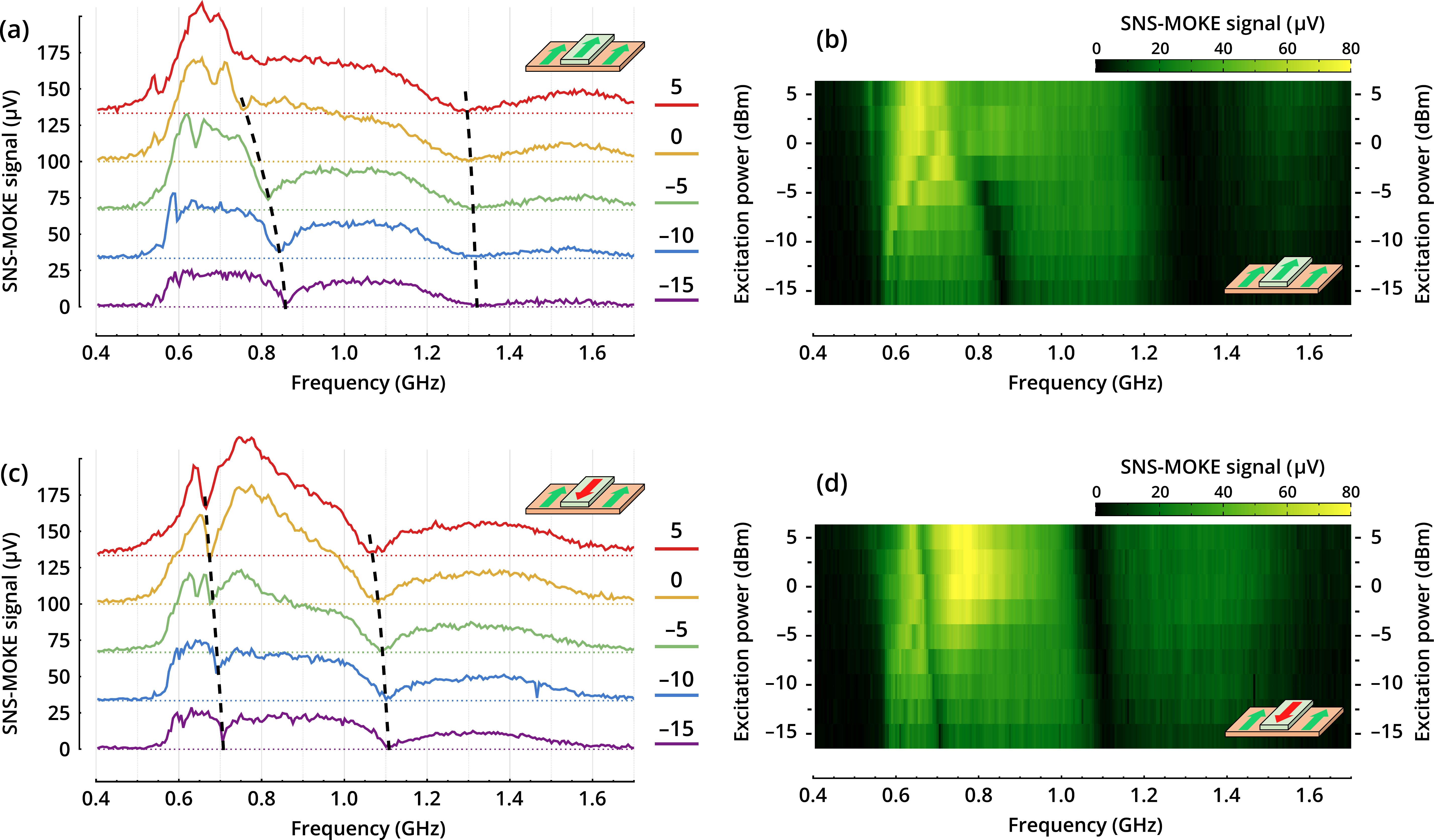}
	\caption{Measured spin-wave transmission spectra for parallel (a,b) and antiparallel (c,d) YIG-CoFeB magnetization alignments are shown for excitation powers between $-15$ and 5 dBm. In (a) and (c), the curves are vertically offset for clarity, and the dashed lines highlight the power-dependent frequency shift of the transmission gaps. The same shifts are visible as the dark contrast in the spectral maps shown in (b) and (d). The insets show the orientations of the YIG and CoFeB magnetizations, with spin waves propagating from left to right.}
	\label{Fig2}
\end{figure*}

In this work, we explore the nonlinear behavior of magnonic Fabry-P\'{e}rot resonators. Using super-Nyquist sampling magneto-optical Kerr effect (SNS-MOKE) microscopy together with micromagnetic simulations, we show that the strong enhancement of spin-wave intensity within the YIG film leads to a frequency downshift of the Fabry-P\'{e}rot resonances as the excitation power increases. These shifts enable the emulation of neuron-like activation behavior and frequency-selective nonlinear spin-wave absorption.  

The sample consists of an 85-nm-thick YIG film with a series of Co$_{40}$Fe$_{40}$B$_{20}$ (CoFeB) nanostripes patterned on top. The YIG film was grown on a (111)-oriented Gd$_3$Ga$_5$O$_{12}$ (GGG) substrate by pulsed laser deposition (PLD) at room temperature and subsequently crystallized by annealing at 750$^\circ$C for 8 hours in 13 mbar of oxygen. The YIG film exhibits a Gilbert damping parameter of $\alpha=5\times10^{-4}$. Electron-beam lithography followed by magnetron sputtering and lift-off was used to fabricate 30 nm thick and 15 $\upmu$m long CoFeB nanostripes with a width range from 350 to 1000 nm, each forming a magnonic Fabry-P\'{e}rot resonator with different spectral characteristics. A 1 nm Ta/3 nm TaO$_x$ spacer between the CoFeB and YIG films blocked exchange while mediating dynamic dipolar coupling between their magnetizations. Spin waves in the YIG film were launched via a 1.5-$\upmu$m-wide Au microwave antenna at excitation powers between $-15$ and $+5$ dBm, with an antenna positioned 25 $\upmu$m from each CoFeB nanostripe (Fig. \ref{Fig1}). Spin-wave transmission through the resonator was measured using SNS-MOKE microscopy\cite{Dreyer2021,Qin2021}. A 3 mT magnetic field was applied parallel to the antenna to establish the Damon-Eshbach geometry.

We complemented the experiments with micromagnetic simulations performed using MuMax3\cite{Vansteenkiste2014}. For the 85-nm-thick YIG film, we used a saturation magnetization $M_{s}=1.2\times10^5$ A/m, an exchange constant $A_\mathrm{ex}=3.5\times10^{-12}$ J/m, and a Gilbert damping parameter $\alpha=5\times10^{-4}$. The CoFeB nanostripe (30 nm thick, 850 nm wide) was modeled using $M_{s}=1.15\times10^6$ A/m, $A_\mathrm{ex}=1.6\times10^{-11}$ J/m, and $\alpha=5\times10^{-3}$. A 5-nm-thick gap separated the nanostripe from the YIG film. The simulation area was discretized into cuboidal cells with 5 nm edge lengths, using a grid of $32768\times2\times24$ cells in the $x$, $y$, and $z$ directions. The 15-$\upmu$m-long CoFeB nanostripe was approximated as infinitely long through one-dimensional periodic boundary conditions along its length ($y$-axis), providing accurate results just behind the resonator \cite{Lutsenko2025}. The quasi-infinite extent of the YIG film along $x$ was modeled using periodic boundary conditions at the left and right edges of the simulation area. To prevent spin-wave circulation caused by periodicity, we implemented identical absorbing Gilbert boundary conditions at both $x$-boundaries. Spin waves were excited in YIG using a continuous-wave Oersted field with a spatial profile corresponding to the 1.5-$\upmu$m-wide microwave antenna placed 25 $\upmu$m from the resonator. The Oersted field was obtained from COMSOL simulations using the Coils module, which reproduced the antenna geometry, position, and applied power. The field, originally defined on an irregular COMSOL mesh, was interpolated onto the regular MuMax3 grid. Simulations were initially run for 35 ns to ensure steady-state conditions. The dynamic magnetization was then sampled 10 $\upmu$m behind the CoFeB nanostripe, with 12 samples per period of the excitation frequency. The spin-wave transmission coefficient of the Fabry-P\'{e}rot resonator was calculated as the ratio of the complex Fourier amplitudes of the dynamic magnetization obtained from simulations with and without the nanostripe present. 

Figure \ref{Fig2} shows the SNS-MOKE signal as a function of frequency, measured 2 $\upmu$m behind the center of an 850-nm-wide CoFeB nanostripe for different excitation powers. Results are presented for both parallel (top) and antiparallel (bottom) orientations of the YIG and CoFeB magnetizations, as illustrated by the schematics in Fig. \ref{Fig2}. In both cases, the YIG magnetization is parallel to the 3 mT applied magnetic field. The SNS-MOKE signal is proportional to the out-of-plane component $m_z$ of the dynamic magnetization and therefore scales with the amplitude of the transmitted spin wave. The spectra exhibit transmission gaps associated with the $n=2$ and $n=3$ magnonic Fabry-P\'{e}rot resonances. As reported in Refs. \citenum{Qin2021b} and \citenum{Lutsenko2025}, these resonances emerge at lower frequencies in the antiparallel configuration, as dictated by differences in the spin-wave dispersion in the two magnetic configurations of the YIG/CoFeB bilayer. Increasing the excitation power causes the transmission gaps to move toward lower frequencies, with shifts of up to 50 MHz between $-15$ dBm and 5 dBm. Similar nonlinear frequency shifts are observed for other stripe widths (supplementary material Fig. S1).  

\begin{figure}[t]
	\centering
    \includegraphics[width=\columnwidth]{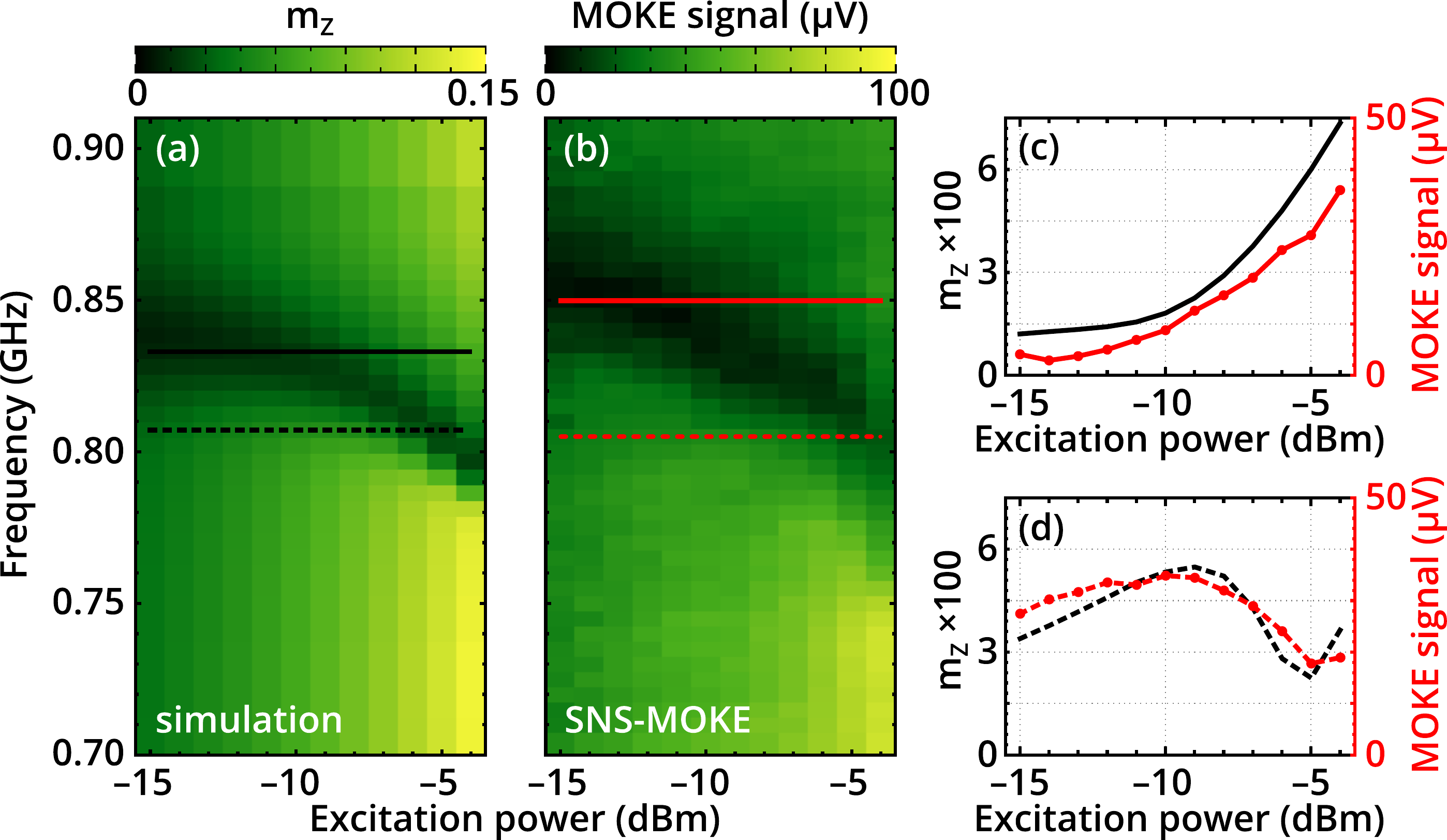}
	\caption{(a) Simulated and (b) measured spin-wave transmission maps for an 850-nm-wide Fabry-P\'{e}rot resonator under excitation powers from $-15$ to $-4$ dBm. The suppression of the spin-wave signal corresponds to the $n=2$ resonance. The simulations reproduce the experimental results with exception of a minor frequency offset, likely arising from small differences between the measured system and its model. (c) Transmitted signal as a function of the excitation power, demonstrating neuron-like threshold activation behavior at $f=0.833$ GHz (simulation) and $f=0.850$ GHz (experiment). (d) Excitation-power dependence of the transmitted signal, showing nonlinear spin-wave suppression at $f=0.807$ GHz (simulation) and $f=0.805$ GHz (experiment).} 
    \label{Fig3}
\end{figure}

To distinguish between nonlinear frequency shifts originating from the Fabry-P\'{e}rot resonator and those intrinsic to the bare YIG film, we performed SNS-MOKE line scans on the same film without the CoFeB nanostripe. These measurements, conducted over frequencies from 0.645 to 1.805 GHz and excitation powers from $-15$ to 5 dBm (supplementary material Fig. S2), reveal that increasing the excitation power increases the spin-wave wavelength. This corresponds to an upward shift of the spin-wave dispersion curve, which is opposite to the experimentally observed downshift of the transmission gaps. Therefore, the latter nonlinearity cannot originate from the intrinsic nonlinear response of the YIG film. Instead, it indicates that nonlinear effects within the Fabry-P\'{e}rot resonator dominate and are sufficiently strong to counteract the nonlinear behavior of propagating spin waves in YIG.    

\begin{figure}[t]
	\centering
	\includegraphics[width=\columnwidth]{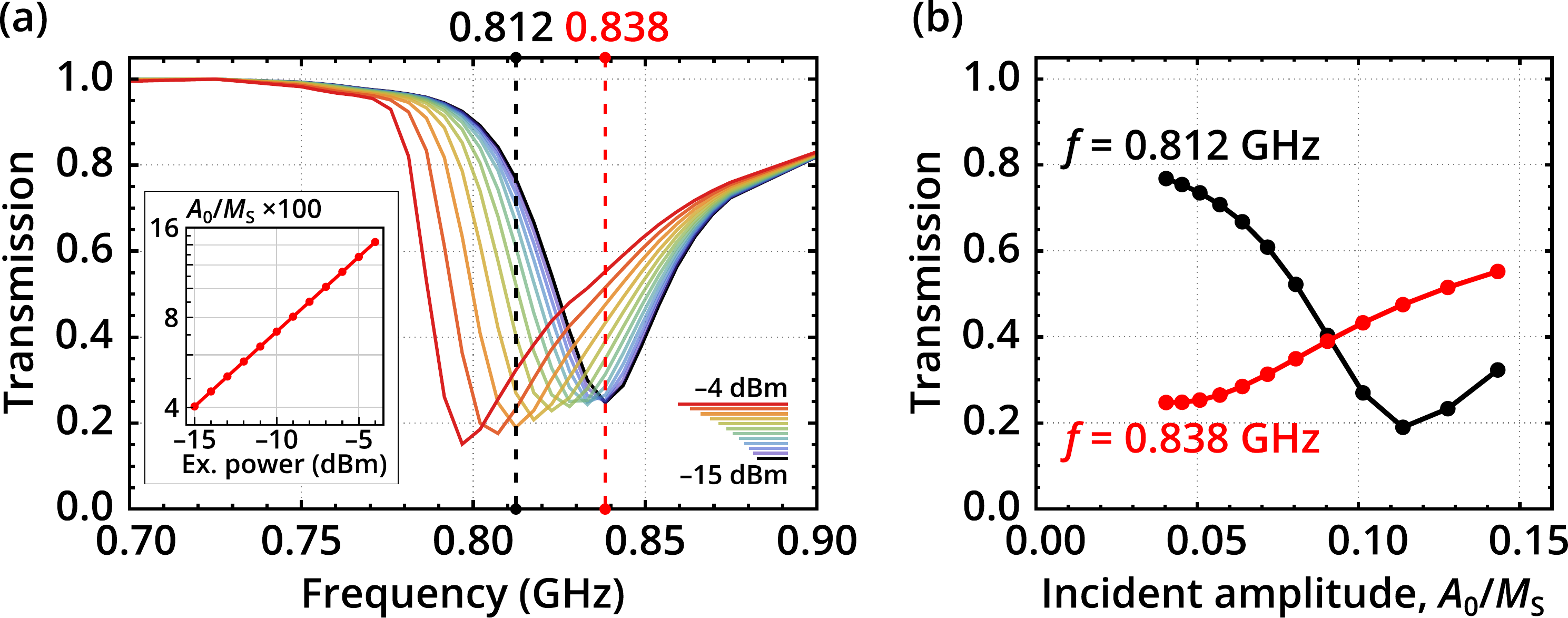}
	\caption{(a) Simulated spin-wave transmission as a function of the incident amplitude $A_0$ for waves crossing an 850-nm-wide resonator, obtained by comparing simulations performed with and without the CoFeB nanostripe. Line color corresponds to the excitation power, ranging from $-15$ to $-4$ dBm. The values of the incident amplitude corresponding to different values of the excitation power (in dBm) are depicted in the inset. (b) Simulated neuron-like threshold activation (red) and nonlinear spin-wave suppression (black) in the isolated magnonic Fabry-P\'{e}rot resonator at $f=0.838$ GHz and $f=0.812$ GHz, respectively.}
    \label{Fig4}
\end{figure}

Micromagnetic simulations reproduce the downward frequency shift of the spin-wave transmission gaps with increasing excitation power, as demonstrated for the $n=2$ resonance in Figs. \ref{Fig3}(a) and \ref{Fig3}(b). Depending on the excitation frequency, this nonlinear shift gives rise to qualitatively different transmission behaviors. As in Ref. \citenum{Fripp2023}, when the excitation frequency lies within the transmission gap in the linear regime (i.e., at low power), increasing the power leads to a nonlinear enhancement of the transmitted spin-wave amplitude (Fig. \ref{Fig3}(c)). This response resembles neuron-like activation functions commonly used in artificial neural networks\cite{Parhi2020}. In contrast, for frequencies just below the transmission gap, higher excitation powers suppress the transmitted spin-wave signal (Fig. \ref{Fig3}(d)). This nonlinear suppression can serve as a magnonic limiter, protecting downstream components from high-power microwave inputs. Both response types can enable nonlinear magnonic logic, where threshold-dependent suppression or enhancement of spin-wave transmission supports logical operations based on amplitude encoding.     

To isolate the nonlinear behavior of a single resonator from the intrinsic nonlinearity of YIG, we performed micromagnetic simulations both with and without the CoFeB nanostripe. Figure \ref{Fig4}(a) shows transmission spectra obtained by calculating the ratio of the complex Fourier amplitudes of the dynamic magnetization in both cases, expressed as a function of the incident wave amplitude $A_0$. The $n=2$ resonance suppresses the transmitted spin-wave signal by approximately 80\%, which is accompanied by a substantial frequency downshift as $A_0$ increases. Figure \ref{Fig4}(b) depicts two distinct transmission characteristics at selected frequencies. At the onset of nonlinear behavior in the resonator, the incident spin-wave amplitude corresponds to only 4\% of the YIG saturation magnetization ($M_s$), and the nonlinearity grows rapidly as $A_0/M_s$ increases to 0.14. In contrast, spin-wave propagation in the bare YIG film remains nearly linear within this simulated excitation range (see inset in Fig. \ref{Fig4}(a)).

\begin{figure}
	\centering
	\includegraphics[width=1.0\columnwidth]{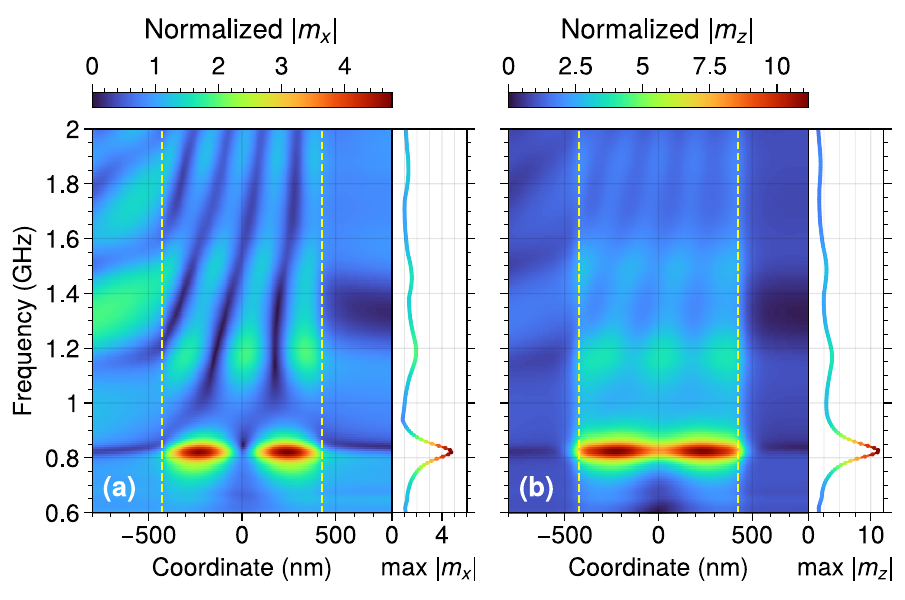}
	\caption{Profiles of the Fourier magnitudes calculated from the simulated in-plane, $|m_x|$, and out-of-plane, $|m_z|$, magnetization components are shown as a function of frequency in the left and right panels, respectively. The $|m_x|$ and $|m_z|$ values for the top surface of the YIG film in a 850-nm-wide magnonic Fabry-P\'{e}rot resonator in the parallel configuration are normalized by their corresponding values obtained for a bare YIG film. The vertical yellow dashed lines mark the edges of the resonator. The line scans show the maximum values of $|m_x|$ and $|m_z|$ within the resonator. Results for the antiparallel configuration as well as a 350-nm-wide resonator are provided in Figs. S3 to S5 of the supplementary material.}
    \label{Fig5}
\end{figure}

The pronounced nonlinearity of the magnonic Fabry-P\'{e}rot resonators at relatively low powers arises from the resonant concentration of the spin-wave energy within the cavity. In contrast to the chiral resonator reported in Ref. \citenum{Fripp2023}, the energy is concentrated in the low-damping YIG film rather than in the ferromagnetic-metal nanostripe. Figure \ref{Fig5} presents frequency-resolved maps of the in-plane and out-of-plane components of the dynamic magnetization in the top layer of the YIG film for the parallel magnetization alignment. Their absolute values are normalized to those obtained from simulations of the bare YIG film without the CoFeB nanostripe. Within the resonator, the spin-wave amplitude is enhanced by about five times for the in-plane component and by more than a factor of eleven for the out-of-plane component compared to the bare YIG film. Consequently, even low-amplitude, nominally linear in bare YIG spin waves produce a nonlinear response when their amplitude is resonantly enhanced inside the nanoscale resonator. 

Our results suggest a route toward magnonic neural networks, where all-to-all connectivity is mediated by interference of linear spin waves, while nonlinearity is selectively introduced at resonator nodes \cite{Kruglyak2021}. The resonant reduction of the nonlinear thresholds suggest that the energy consumption in such devices will be substantially reduced in comparison to non-resonant neural networks \cite{Papp2021}. Alternatively, nonlinear magnonic Fabry-P\'{e}rot resonators could be used within neuromorphic computing architectures that utilize time-delay multiplexing\cite{Watt2021}, enabling both their reservoir computing and recurrent neural network functionalities. 

In summary, we have investigated the power-dependent response of magnonic Fabry-P\'{e}rot resonators. Both experiments and simulations reveal a systematic frequency downshift of the spin-wave transmission gaps with increasing excitation power, enabling the realization of neuron-like activation functions and frequency-selective nonlinear spin-wave absorption. Compared to a bare YIG film, the threshold power for nonlinear resonator behavior is reduced, highlighting the potential of these nanoscale resonators as low-power nonlinear elements for neuromorphic magnonic computing networks.\newline

This project has received funding from the European Union’s Horizon Europe research and innovation program under Grant Agreement No. 101070347-MANNGA. However, views and opinions expressed are those of the authors only and do not necessarily reflect those of the European Union or the European Health and Digital Executive Agency (HADEA). Neither the European Union nor the granting authority can be held responsible for them. The project also received funding from the Research Council of Finland (Grant No. 357211) and the UK Research and Innovation (UKRI) under the UK government’s Horizon Europe funding guarantee (Grant No. 10039217) as part of the Horizon
Europe (HORIZON-CL4-2021-DIGITAL-EMERGING-01) under
Grant Agreement No. 101070347.\\ 

\noindent \textbf{AUTHOR DECLARATIONS}

\noindent \textbf{Conflict of Interest}

The authors have no conflicts to disclose.\\
	
\noindent \textbf{Author Contributions}

A.L. performed the experiments. A.L. and K.G.F. conducted the micromagnetic simulations. L.F. assisted in the SNS-MOKE measurements. A.V.S., V.V.K., and S.v.D supervised the work. All authors discussed the results and contributed to writing of the manuscript.\\ 
	
\noindent \textbf{DATA AVAILABILITY}

The data that support the findings of this study are available from the corresponding authors upon reasonable request.\\

\noindent \textbf{REFERENCES}
\bibliography{references}

\end{document}